\documentclass[%
 reprint,
 amsmath,amssymb,
 aps,
]{revtex4-2}

\pdfoutput=1
\UseRawInputEncoding
\usepackage{graphicx}
\usepackage{dcolumn}
\usepackage{bm}

\begin{document}


\title{A new type-II lepidocrocite-type TiO$_{2}$/GaSe heterostructure: Electronic and optical properties, bandgap engineering, interaction with ultrafast laser pulses}

\author{Yilin Zhao\textsuperscript{1},Hong Zhang\textsuperscript{1,2}}
 \email{hongzhang@scu.edu.cn}
\affiliation{%
 \textsuperscript{1}College of Physics, Sichuan University, Chengdu 610065, PR China\\
}%
\author{Xinlun Cheng\textsuperscript{2}}%
\affiliation{%
 \textsuperscript{2}Key Laboratory of High Energy Density Physics and Technology of Ministry of Education, Sichuan University, Chengdu 610065, PR China\\
}%





\begin{abstract}
Recently, van der Waals heterostructure has attracted interest both theoretically and experimentally for their potential applications in photoelectronic devices, photovoltaic devices, plasmonic devices and photocatalysis. Inspired by this, we design a lepidocrocite-type TiO$_{2}$/GaSe heterostructure. Via first-principles simulations, we show that such a heterostructure is a direct bandgap semiconductor with a strong and broad optical absorption, ranging from visible light to UV region, exhibiting its potential application in photoelectronic and photovoltaic devices. With the planar-averaged electron density difference and Bader charge analysis, the heterostructure shows a strong capacity of enhancing the charge redistribution especially at the interface, prolonging the lifetime of excitons, and hence improving photocatalytic performance. By applying biaxial strain and interlayer coupling, the heterostructure exhibits a direct-indirect bandgap transition and shows a potential for mechanical sensors due to the smooth and linear variation of bandgaps. Furthermore, our result indicates that a lower interlayer distance leads to a stronger charge redistribution. The calculation of irradiating ultrafast on the heterostructure further reveals a semiconductor-metal transition for the heterostructure. Moreover, we find an enhanced induced plasmonic current in the heterostructure under both x-polarized and z-polarized laser, which is beneficial to plasmonic devices designs. Our research provides valuable insight in applying the lepidocrocite-type TiO$_{2}$/GaSe heterostructure in photoelectronic, photovoltaic, photocatalytic, mechanical sensing and plasmonic realms. 
\end{abstract}

\maketitle


\section{\label{sec:level1}INTRODUCTION}

Recent years, two-dimensional (2D) materials have come into notice for their unusual properties, which mainly arise from the quantum confinement and the original structure characteristics. Assembling 2D materials into van der Waals heterostructures is one of the most efficient and achievable ways to take advantage of all kinds of 2d materials, yielding a range of applications in tunneling devices\cite{RN1,RN2,RN3}, optoelectronic devices\cite{RN4,RN5,RN6,RN7}, photovoltaic devices\cite{RN8,RN9,RN10,RN11,RN12}, plasmonic devices\cite{RN13,RN14,RN15,RN16}, light-emitting diodes\cite{RN12,RN17,RN18}, etcetera.

Gallium selenide (GaSe) is a layered semiconducting material, composed of four sublayers stacking in the sequence of Se-Ga-Ga-Se. The monolayer GaSe sheets have been successfully synthesized in experiment\cite{RN19,RN20,RN21,RN22}, which is widely used in optoelectronics\cite{RN23,RN24}, nonlinear optics\cite{RN25}, terahertz experiments\cite{RN26}, solar energy conversion\cite{RN27}, field-effect transistors (FETs)\cite{RN21}, and so on.

Titanium dioxide (TiO$_2$), due to its unique properties, such as environment-friendly, photo and chemical stability and low production cost\cite{RN28,RN29}, is one of the most prominent semiconductors with a large range of applications in photocatalysis\cite{RN30,RN31,RN32}, photovoltaic cells\cite{RN33,RN34,RN35,RN36}, and so forth. However, the large bandgap limits its light absorption window to ultraviolet (UV) range. Besides, the fast recombination of photo generated electron and hole pairs makes TiO$_2$ less efficient in photocatalysis\cite{RN37}. Compared with other configurations of TiO$_2$, a lepidocrocite-type TiO$_2$ has been experimentally synthesized with a larger bandgap (about 3.80eV) and stronger redox power\cite{RN38,RN39,RN40}. Theoretical and experimental efforts have been carried out to seek ways to improve the photocatalytic activities of lepidocrocite-type TiO$_2$\cite{RN30,RN31,RN41,RN42}. Coupling TiO$_2$ with other 2D monolayers into heterostructure is one of the effective approaches. However, to the best of our knowledge, among these TiO$_2$ heterostructures, little is known about the electronic properties of lepidocrocite-type TiO$_2$/GaSe heterostructure.

In this work, we theoretically studied the structural, electronic and optical properties of the lepidocrocite-type TiO$_2$/GaSe heterostructure. The direct bandgap, type-II band alignment, strong and wide optical absorption suggest its potential application in photovoltaic and photoelectronic devices. To further explore the charge transference of the heterostructure, the planar-averaged electron density difference and Bader charge analysis was carried out to reveal its merit in intensifying the charge redistribution and serving as photocatalyzer. We also investigated how the biaxial strain and interlayer distance can be applied to tune the band structure of the heterostructure, increasing the possibility for a wider range of application. Finally, we studied the ultrafast laser acting on the heterostructure, further revealing its potential application in plasmonic devices.

The paper is organized as follows. In Sec.\ref{sec:2}, we describe the computational details employed in this paper. In Sec.\ref{sec:3.1}, we investigate the band structures, electronic properties of GaSe and TiO$_2$ monolayers. In Sec.\ref{sec:3.2} and \ref{sec:3.3}, we study the electronic and optical properties of the TiO$_2$/GaSe heterostructure. In Sec.\ref{sec:3.4}, we explore the approaches to modulate the band structure with biaxial strain and interlayer coupling. In Sec.\ref{sec:3.5}, we have a deep study on ultrafast lasers acting on the heterostructure. In Sec.\ref{sec:4}, we summarize our results.

\section{COMPUTATIONAL DETAILS}
\label{sec:2}

Our calculations were mainly based on the first-principles density functional theory (DFT). The projector-augmented wave (PAW) method was implanted in the Vienna ab initio simulation package (VASP)\cite{RN43,RN44}. The Perdew-Burke-Ernzerhof (PBE) version of the generalized gradient approximation (GGA) was employed\cite{RN45}. The Van der Waals (vdW) interaction between TiO$_2$ and GaSe layers was corrected with the DFT-D3 method of Grimme\cite{RN46,RN47}. As we all know, the DFT method is not quite accurate when dealing with transition-metals systems with localized electrons (d or f). As a result, the DFT+U method was applied to describe the localized Ti 3d electrons in the following equation:
\begin{equation}
E_{U}=\frac{U}{2} \sum_{I, \sigma} \operatorname{Tr}\left[\boldsymbol{n}^{I \sigma}\left(1-\boldsymbol{n}^{l \sigma}\right)\right]
\end{equation}
\begin{equation}
E_{P B E+U}-E_{P B E}+\frac{U-J}{2} \sum_{\sigma} \operatorname{Tr}\left[\rho^{\sigma}-\rho^{\sigma} \rho^{\sigma}\right]
\end{equation}
Where $n^{I\sigma}$ denotes the occupation of the relevant localized manifold at site I with spin $\sigma$, U denotes the effective on-site Coulomb interaction\cite{RN48}. With a self-consistent DFT+U approach\cite{RN49,RN50} performing on Quantum Espresso (QE), an open source first principles code, the value of U was determined to be 4.75eV for the Ti 3d electrons. The vacuum region along z direction was set to be 20Å and the energy cutoff was set to be 700eV through all calculations. For the structure relaxations, the k-points sampling with $\Gamma$ centered Monkhorst–Pack Scheme was chosen to be $8\times13\times1$,$15\times15\times1$,$13\times16\times1$ for unit cells of the heterostructure, GaSe and TiO$_2$ respectively. All structures are fully relaxed until the forces were less than 0.01eV/Å and the energy tolerances were smaller than $1\times10^-6$eV/atom. For the study of optical properties, the optical absorption coefficient $\alpha(\omega)$ was derived from the following formula,
\begin{equation}
\alpha(\omega)=\sqrt{2} \omega\left[\sqrt{\varepsilon_{1}^{2}(\omega)+\varepsilon_{2}^{2}(\omega)}-\varepsilon_{1}(\omega)\right]^{\frac{1}{2}}
\end{equation}
Where $\epsilon_1(\omega)$ and $\epsilon_2(\omega)$ are the real and imaginary part of the complex dielectric function. For the calculation of the planar-averaged charge density, a fully relaxed 2x2 unit cells of TiO$_2$/GaSe heterostructure was chosen, and the k-points sampling was set to be $4\times7\times1$. Moreover, the planar-averaged electron density difference $\Delta\rho(z)$ is defined as:
\begin{align}
\Delta \rho(\mathrm{z})&=\int_{\Sigma(z)}\left[\rho\left(T i O_{2} / \text { GaSe }\right)-\rho\left(T i O_{2}\right)-\rho(\text { GaSe })\right] \notag \\
&=\sum_{i j} \Delta \rho_{i j} \Delta x_{i} \Delta x_{j}
\end{align}
In which, $\rho(TiO_2/GaSe)$, $\rho(TiO_2)$, $\rho(GaSe)$ are the electron density of TiO$_2$/GaSe heterostructure, TiO$_2$, GaSe, respectively. In addition, the interaction between the heterostructure and laser were calculated with the real-space and real-time TDDFT code OCTOPUS\cite{RN51}. To calculate the nonlinear optical response of the system, we solved the time-dependent Schrodinger equation:
\begin{gather}
-\mathrm{i} \frac{\partial}{\partial \mathrm{t}} \Psi_{i}(\boldsymbol{r}, t)=\Psi_{i}(\boldsymbol{r}, t)\times \notag \\
\left[-\frac{\nabla^{2}}{2}+\upsilon_{\text {ext }}(\boldsymbol{r}, t)+\upsilon_{\text {Hartree }}(\boldsymbol{r}, t) +\upsilon_{x c}(\boldsymbol{r}, t)+\upsilon_{\text {laser }}(\boldsymbol{r}, t)\right] 
\end{gather}
Where $\frac{\nabla^{2}}{2}$ denotes the kinetic energy, $\upsilon_{\text {ext }}(\boldsymbol{r}, t)$ is the external potential, $\upsilon_{\text {Hartree }}$ represents the electron-electron interaction, $\upsilon_{x c}(\boldsymbol{r}, t)$ describes the exchange-correlation potential, and $\upsilon_{\text {laser }}(\boldsymbol{r}, t)$ is the time-dependent electromagnetic laser field. The initial state is given by solving the ground-state Kohn-Sham equations. Then, the Kohn-Sham orbitals propagate as:
\begin{equation}
\Psi_{i}(\boldsymbol{r}, t+\Delta t)=e^{-i \int_{t}^{t+\Delta t} d \tau \hat{H}_{K S}(r, t)} \Psi_{i}(\boldsymbol{r}, t)
\end{equation}

\begin{figure}[htbp]
\includegraphics[scale=0.3]{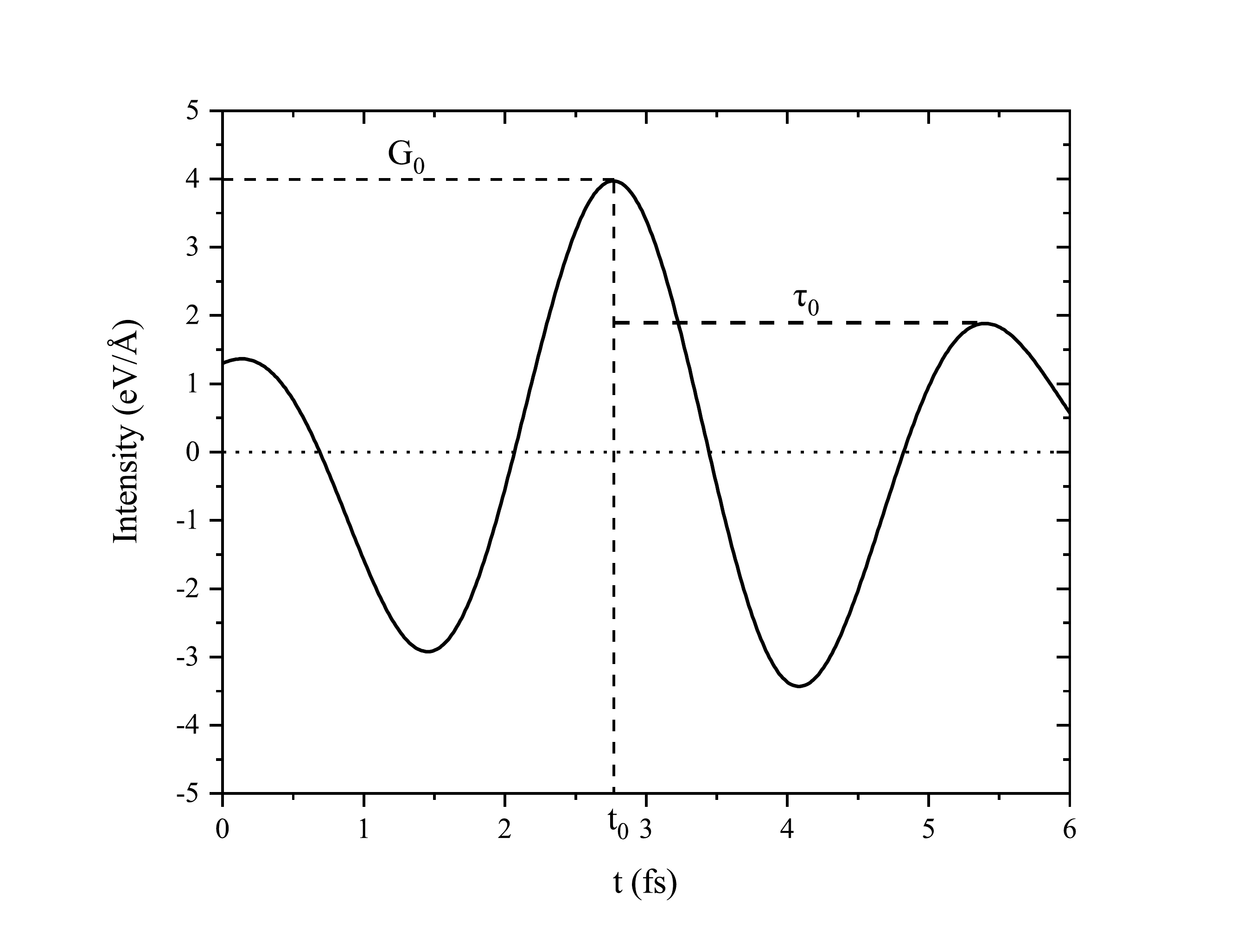}
\caption{\label{fig:1} Illustration of Gaussian laser oscillogram}
\end{figure}

As shown in Fig.\ref{fig:1}, the laser was simulated with the Gaussian wave packet, which is defined as:
\begin{equation}
f(x) \cos [\omega t+\phi(t)] g(t)
\end{equation}
\begin{equation}
g(t)=G_{0} \exp \left[-\left(t-t_{0}\right)^{2} / 2 \tau_{0}^{2}\right]
\end{equation}
In which f(x) denotes the laser polarization direction, $\omega$ represents the circular frequency, $\phi(t)$ defines the initial phase, $G_0$ is the amplitude of g(t) in units of eV/Å (1eV/Å is equal to $1.327x1013W/cm^2$, according to $I=\frac{1}{2}ecE^2$, where e is the dielectric function, c is the light velocity in vacuum), $t_0$ determines where g(t) centers around, and $\tau_0$ is related to the full width at half maximum (FWHM). The elements are described by the PseudoDojo Potentials\cite{RN52}. The generalized gradient approximation (GGA) expressed by Perdew-Burke-Ernzerhof (PBE) functional is used both for the ground-state and excited-state calculations. The simulation box is defined by adding a sphere of radius 8Å around each atom with a uniform mesh grid of 0.3Å. For the time evolution, the time-step is set to be $0.005{\hbar}eV^{-1}$(around 0.0033fs).

\section{RESULTS AND DISCUSSION}
\label{sec:3}
    \subsection{Structures and electronic properties of GaSe and TiO$_2$ monolayer}
    \label{sec:3.1}
\begin{figure}[htbp]
\centering
\includegraphics[scale=0.25]{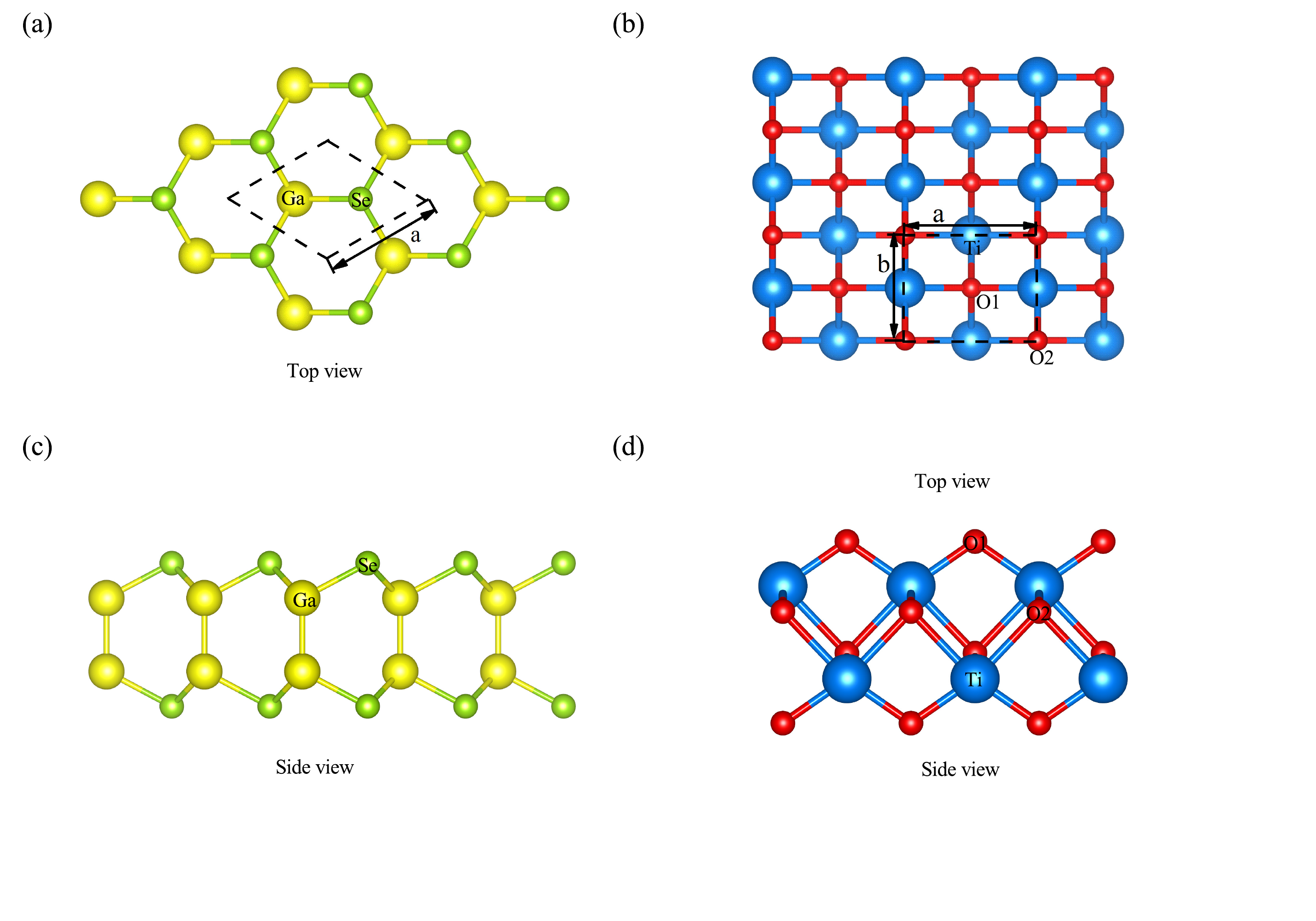}
\caption{\label{fig:2} (a)Top and (c)side view of GaSe monolayer. Yellow, green stand for gallium, selenium, respectively. (b)Top and (d)side view of lepidocrocite-type TiO$_2$ monolayer. Blue, red stands for titanium and oxygen. }
\end{figure}

Before constructing the heterostructure, we first analyzed geometric properties of the GaSe and TiO$_2$ monolayers. For GaSe monolayer, as shown in Fig.\ref{fig:2}, the optimized lattice parameters are a=b=3.81 Å; the Ga-Se bond length $d_{Ga-Se}$ and the Ga-Ga bond length $d_{Ga-Ga}$ are 2.40Å and 2.497Å, respectively, which is in good agreement with previous studies (a=b=3.82Å, $d_{Ga-Se}$ =2.501Å, $d_{Ga-Ga}$=2.470Å)\cite{RN53}. For lepidocrocite-type TiO$_2$ monolayer, we can see from Fig.\ref{fig:2}(b, d) that the lattice parameters are a=3.83 Å, b=3.05 Å. There are two inequivalent O atoms that are 2-fold (O1) and 4-hold (O$_2$) coordinated to Ti atoms. The Ti-O1 bond length $d_{Ti-O1}$is 1.860 Å; The Ti-O$_2$ bond length $d_{Ti-O_2}$ is 2.014 Å; The Ti to O$_2$ in the next unit cell bond length $d_{Ti-O2}^{'}$ is 2.212 Å, which also consists with the values reported by other research groups (a=3.73 Å, b=3.03 Å, $d_{Ti-O1}$=1.82 Å, $d_{Ti-O2}$=2.23 Å, $d_{Ti-O2}^{'}$=1.96 Å)\cite{RN54}.

Meanwhile, we calculated the charge densities of the primitive cells of GaSe and TiO$_2$ monolayers. The Bader charge analysis suggests that 0.66 electron transfers from each Ga atom to per Se atom; each Ti atom lost 2.34 electrons and each O atom at position 1 and 2 gains 1.07, 1.17 of the electrons respectively.

\begin{figure}[htbp]
\centering
\includegraphics[scale=0.5]{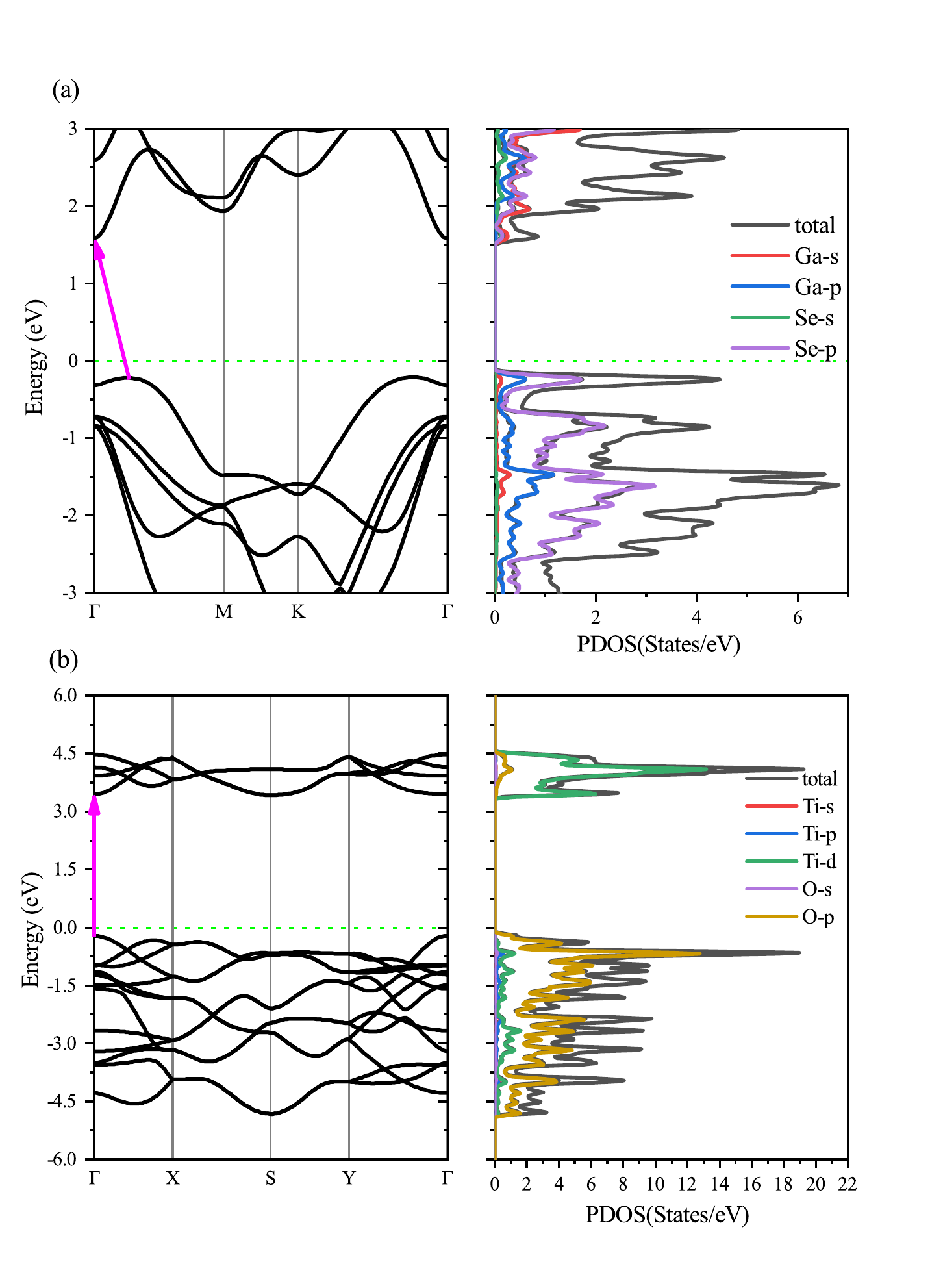}
\caption{\label{fig:3} The band structure and projected density of states (PDOS) of (a) GaSe monolayer and (b) lepidocrocite-type TiO$_2$ monolayer primitive cell. }
\end{figure}

In Fig.\ref{fig:3}(a-b), we further computed the band structures and projected density of states (PDOS) of the primitive cells of GaSe and TiO$_2$ monolayers. Fig.\ref{fig:3} shows that GaSe possess an indirect band-gap of 1.796eV, which well agrees with the value of previous study (1.82eV)\cite{RN55}. The conduction band minimum (CBM) locates at the $\Gamma$ point, which is composed of 4s states of Ga atom and 4s, 4p states of Se atom. The valence band maximum (VBM) lies on the $\Gamma$M line in the Brillouin zone (BZ), which mainly contains the 4s, 4p states of Ga atom and 4p states of Se atom. On the contrary, as shown in Fig.\ref{fig:3}(b), the TiO$_2$ monolayer has a direct band gap of 3.633eV, locating at $\Gamma$ point in the BZ, which is consistent with the experiment value of 3.80eV2 due to the application of DFT+U method. The CBM is contributed by 3d states of Ti atom and 2p states of O atom, while VBM consists of 2p states of O atom.

\subsection{Structures and electronic properties of TiO$_2$/ GaSe heterostructure}
\label{sec:3.2}
\begin{figure}[htbp]
\centering
\includegraphics[scale=0.25]{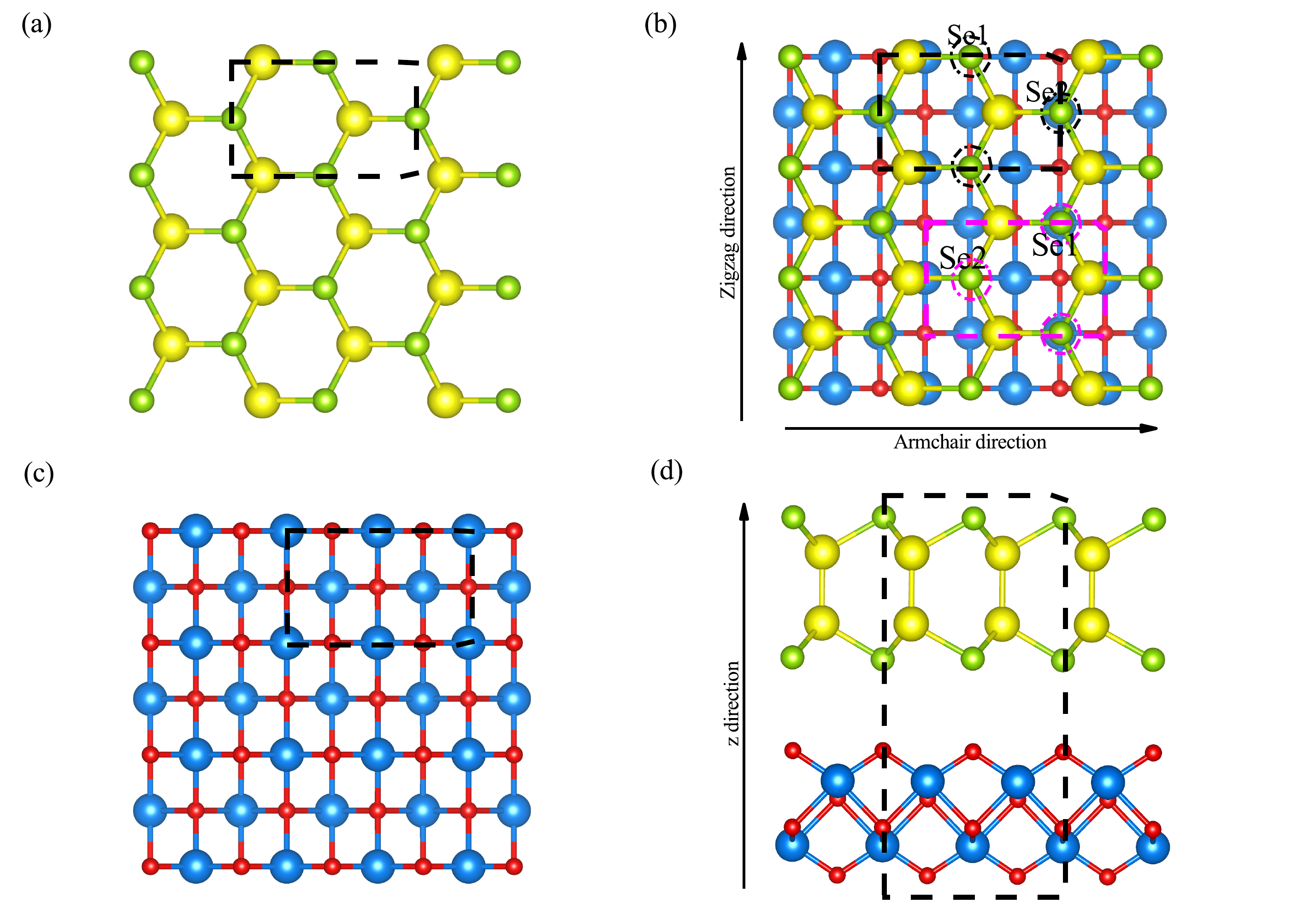}
\caption{\label{fig:4} Top view of (a) GaSe and (c) lepidocrocite-type TiO$_2$ monolayer. (b)Top and (d) side view of the lepidocrocite-type TiO$_2$/GaSe heterostructure. The black dash line rectangles stand for the unit cells of each structure. The vacuum layer is not plotted.}
\end{figure}

To start with, we cleaved a $1\times1$ rectangular GaSe (001) slab and a $2\times1$ TiO$_2$ slab. The van der Waals heterostructure was constructed by stacking the GaSe (001) monolayer on the top of the TiO$_2$ (001) monolayer. The lattice mismatch is less than $4.5\%$. As shown in Fig.\ref{fig:4}(b, d), the optimized lattice parameters of the heterostructure are a=6.22 Å, b=3.82 Å, and the inter layer distance between GaSe and TiO$_2$ monolayer is 2.87 Å. Unlike the GaSe and TiO$_2$ monolayer, the heterostructure is monoclinic, with the space group of Pm. In the unit cell, the Se1 atom is on the top of the O atom and Se2 is on the top of Ti atom; or equivalently, Se1 atom is on the top of Ti atom and Se2 is on the top of O atom.
In order to analyze the stability of the heterostructure, we calculated the binding energy $E_b$, which is defined in the formula:
\begin{equation}
E_{b}=E_{\text {heterostructure}}-E_{\text {Gase}}-E_{\text {Tio}_{2}}
\end{equation}
In which, $E_{heterostructure}$, $E_{GaSe}$, $E_{TiO_2}$are the total energy of the heterostructure, GaSe monolayer and TiO$_2$ monolayer, respectively. By this definition, a more negative $E_b$ indicates a more stable structure. The value of $E_b$ is calculated as -2.491eV, indicating a good stability of the heterostructure and a strong interaction of the two monolayers.

\begin{figure*}[htbp]
\centering
\includegraphics[scale=0.5]{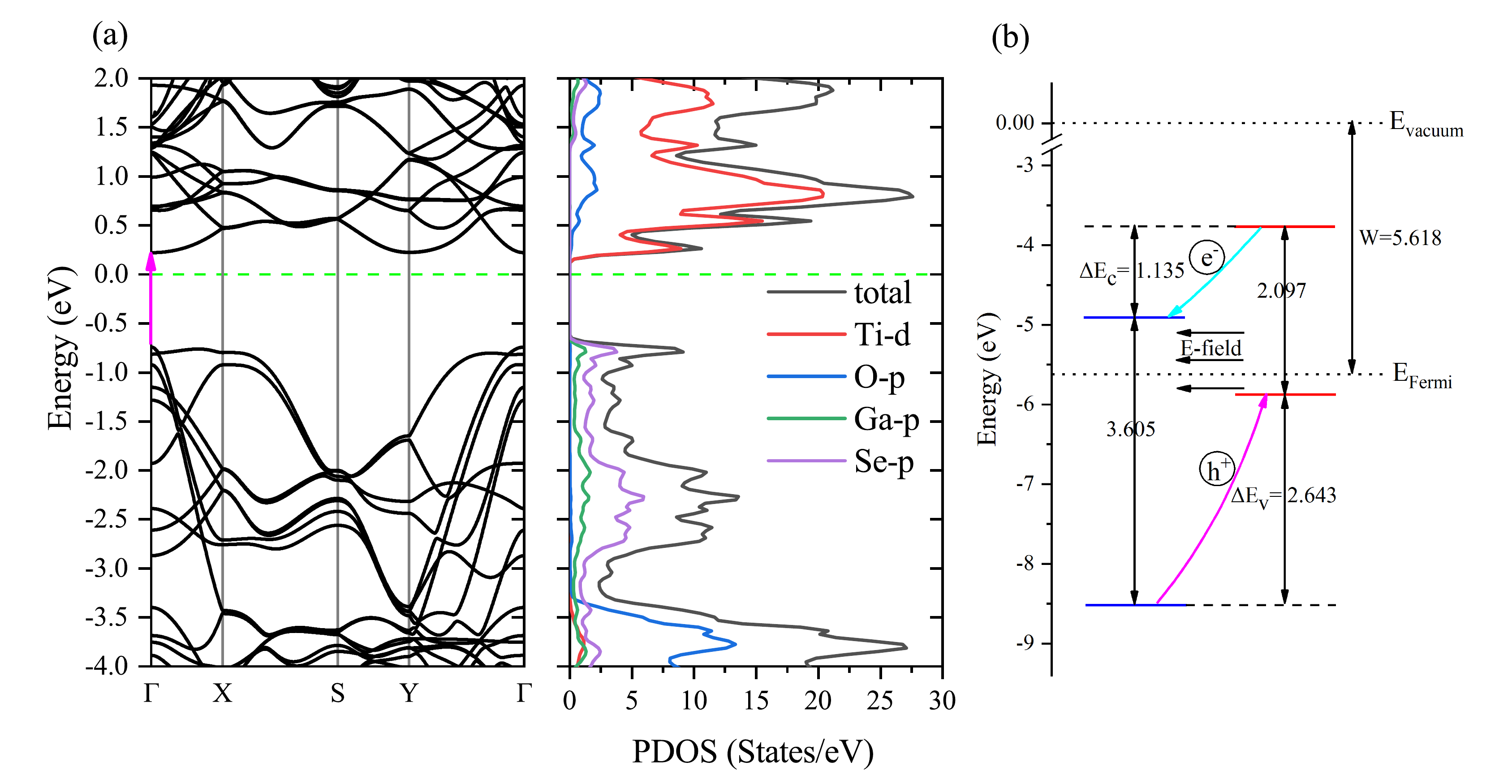}
\caption{\label{fig:5}(a) The band structure and projected density of states (PDOS) of TiO$_2$/ GaSe heterostructure. (b) Illustration of the band alignment of the TiO$_2$/GaSe heterostructure with respect to the vacuum level.}
\end{figure*}

The band structure and partial density of states of TiO$_2$/GaSe heterostructure are shown in Fig.\ref{fig:5}(a). As we can see, TiO$_2$/GaSe heterostructure is a semiconductor with a direct band-gap of 0.963eV; the CBM and VBM both lie at the $\Gamma$ point, which indicates that the lowest-energy electron-hole pairs can be spontaneously separated, making TiO$_2$/GaSe heterostructure suitable for applications in photovoltaic devices. The CBM are mainly composed of the 3d states of Ti atom, while the VBM is contributed by the 4p states of Ga, Se atoms. It is also clear that the CBM of TiO$_2$ is lower than the CBM of GaSe; the VBM of TiO$_2$ is also lower than the VBM of GaSe, which exhibits a type-II band alignment. For further clarification, the band alignment with reference to the vacuum level ($E_vacuum$) is shown in Fig.\ref{fig:5}(b). The band offsets of the heterostructure are ${\Delta}E_c$=1.135eV and ${\Delta}E_v$=2.643eV. For comparison, the natural band offsets of individual TiO$_2$ and GaSe monolayer are ${\Delta}E_c$=0.931eV and ${\Delta}E_v$=2.773eV. The difference in band offsets is mainly due to the charge redistribution when forming heterostructure. When light irradiating, the photoexcited electrons can transfer from the valence band (VB) to the conduction band (CB) both in the TiO$_2$ and GaSe monolayers. At the same time, photoexcited holes are left in the VBs. Due to the conduction band offset ${\Delta}E_c$, the photo-excited electrons on GaSe CB tend to flow to the CB of TiO$_2$ nano sheet. Meanwhile, driven by the valence band offset ${\Delta}E_v$, the left holes shift from the VB of TiO$_2$ surface to the VB of GaSe surface. This kind of charge redistribution benefits the separation of photoexcited electrons and holes, prolonging the lifetime of excitons, which accelerates the oxidation and redox reactions on the surface and improves the efficiency of photocatalytic activity. 

\begin{figure*}[htbp]
\centering
\includegraphics[scale=0.5]{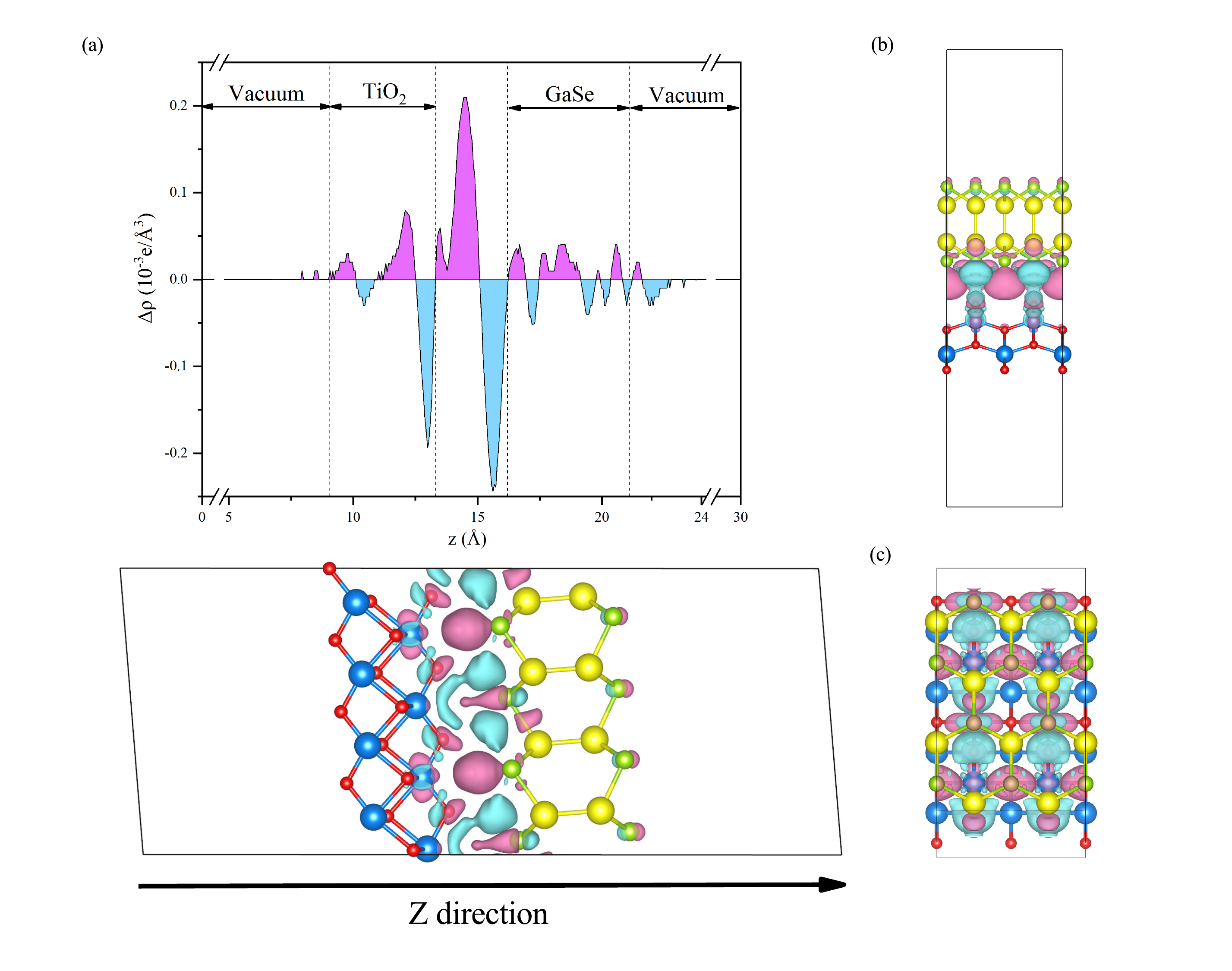}
\caption{\label{fig:6}(a) The planar-averaged electron density difference $\Delta\rho(z)$ along z direction of the TiO$_2$/GaSe heterostructure. (b) Side and (c) top view of the charge density difference of the TiO$_2$/GaSe heterostructure. The purple and cyan denote electron accumulation and depletion regions, respectively. }
\end{figure*}

To quantitatively investigate the charge redistribution activity, we calculated the z-direction planar-averaged electron density difference $\Delta\rho(z)$ of the TiO$_2$/GaSe heterostructure, as shown in Fig.\ref{fig:6}(a). The charge accumulation region is depicted in purple, while the charge depletion region is shown in cyan. The planar-averaged electron density indicates that the charge redistribution mainly occurs at the interface of the heterostructure and the electrons tend to flow from GaSe surface to TiO$_2$ surface, which is in accordance with the band alignment analysis. To quantitively gain an insight of the charge transference, we carried out the Bader charge analysis of the TiO$_2$/GaSe heterostructure. The result suggests that 0.099 electrons transfers from GaSe layer to the TiO$_2$ layer. The net charge accumulation leads to the formation of a built-in electric field at the interface of the TiO$_2$/GaSe heterostructure. The direction of the electric field is from GaSe monolayer to the TiO$_2$ monolayer, which in turn hinders the flow of the electrons and holes, and the system will finally achieve equilibrium as the diffusion force is balanced with the built-in electric field.

\begin{figure*}[htbp]
\centering
\includegraphics[scale=0.5]{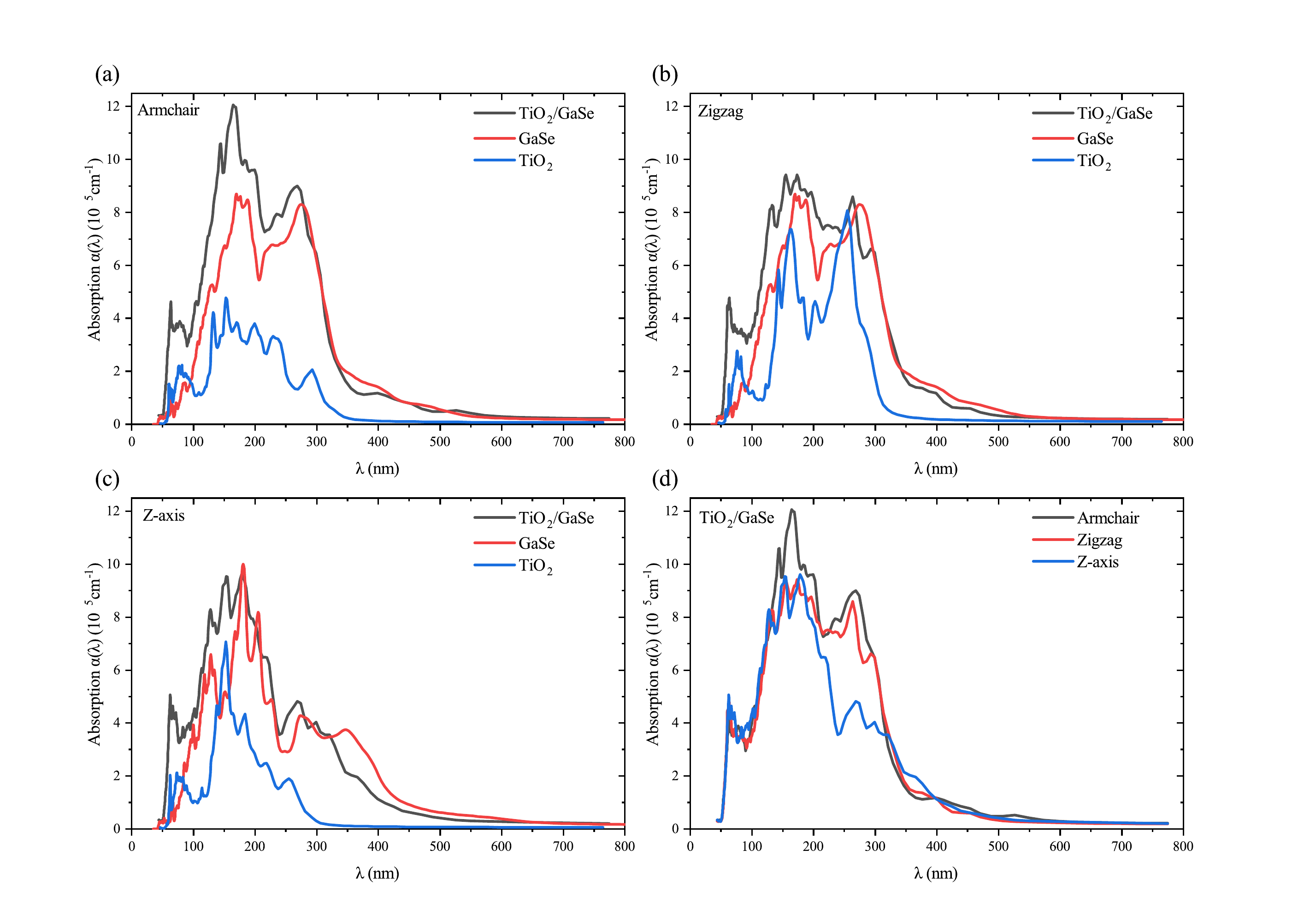}
\caption{\label{fig:7} The optical absorption coefficient $\alpha(\lambda)$ of TiO$_2$/GaSe heterostructure, GaSe monolayer, TiO$_2$ monolayer along (a) armchair direction, (b) zigzag direction, (c) z-axis. (d) Comparison of the absorption coefficient of the TiO$_2$/GaSe heterostructure along the armchair, zigzag and z direction. }
\end{figure*}

\subsection{Optical properties of TiO$_2$/GaSe heterostructure and TiO$_2$, GaSe monolayers}
\label{sec:3.3}
To further explore the photoelectronic properties of TiO$_2$/GaSe, we calculated the optical absorption coefficient $\alpha(\omega)$ of the heterostructure, GaSe slab and TiO$_2$ slab along the armchair, zigzag and z directions, as exhibited in Fig.\ref{fig:7}(a-d). The absorption range of the heterostructure extends broadly from visible light to the ultraviolet (UV) region, and the absorption intensity reaches an order of $10^5$.  Compared with TiO$_2$ and GaSe monolayer, the heterostructure’ s absorption behavior is clearly enhanced in UV region, which indicates that TiO$_2$/GaSe heterostructure can be used for developing electro-photonic detectors. The absorption intensity along armchair direction is stronger than those along the zigzag and z-axis directions in 120~220nm, and the absorption intensity along z axis is weaker than those along the armchair and zigzag directions in 220~320nm. The anisotropic behavior makes TiO$_2$ heterostructure suitable for serving as polarized optical sensors. 

\subsection{Tuning band structures of heterostructure with biaxial strain and interlayer coupling}
\label{sec:3.4}
\begin{figure*}[hbtp]
\centering
\includegraphics[scale=0.5]{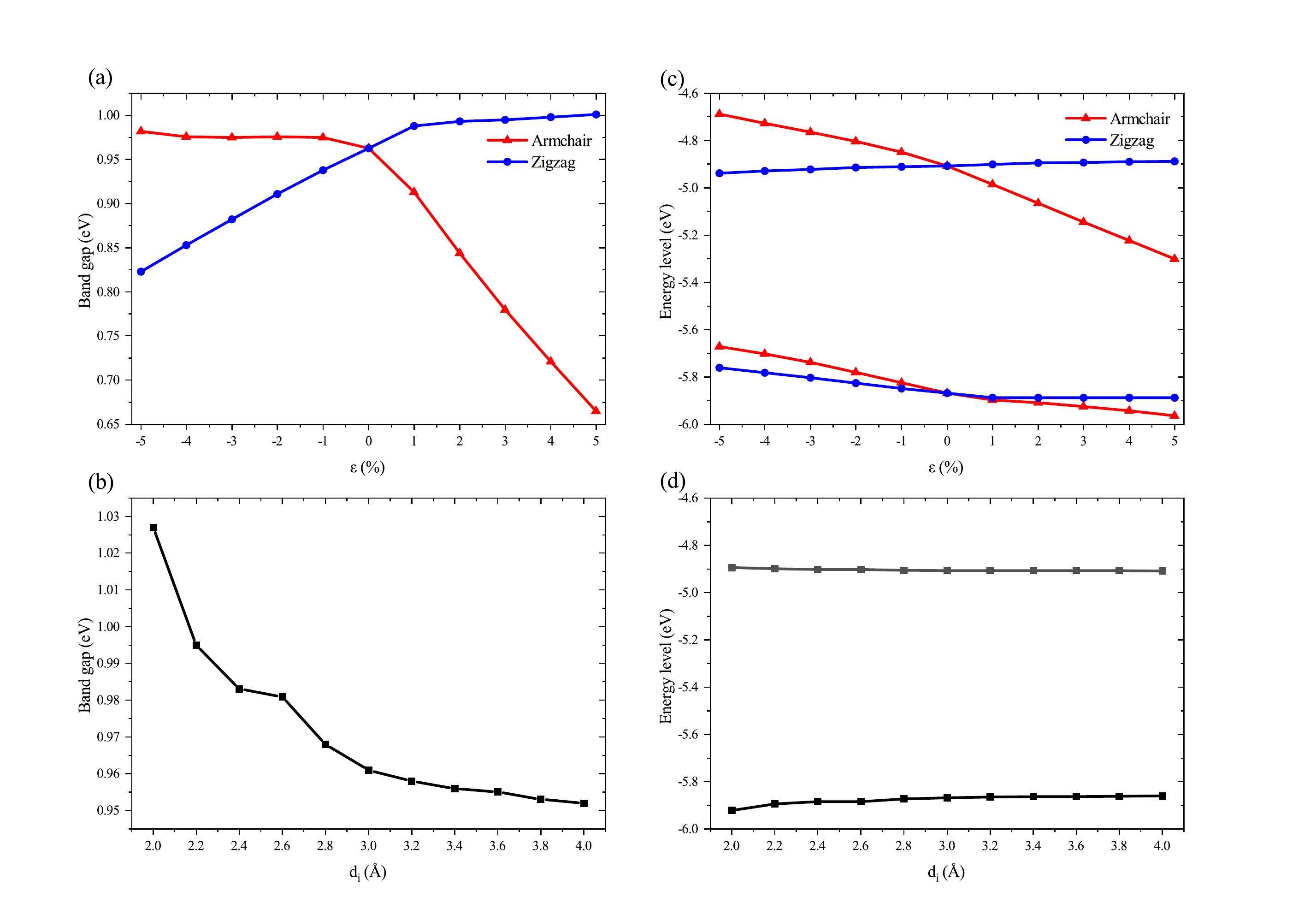}
\caption{\label{fig:8} (a) Bandgap, (b) CBM and VBM of the TiO$_2$/GaSe heterostructure as a function of strain $\varepsilon$ along the armchair and zigzag direction, respectively. (b) Bandgap, (d)CBM and VBM of the TiO$_2$/GaSe heterostructure as a function of interlayer distance di.}
\end{figure*}

For the application of 2D materials in real systems, the modulation of electronic structure has always been an important question. It has been clarified in many experimental and theoretical studies that the mechanical strain and interlayer coupling play an important role in modulating band structures and electronic properties of materials\cite{RN56,RN57,RN58,RN59,RN60}. In this case, we first studied the change in the bandgaps and band edges of the TiO$_2$/GaSe heterostructure when applying biaxial tensile and compressive strains in the xy-plane. The biaxial strain along x and y direction can be characterized by the degree in which the lattice constant differs from its optimized value: $\varepsilon=(a-a_0)/a_0$, where $a_0$ denotes the optimized lattice parameter. The tensile and compressive strain is characterized by the positive and negative value of $\varepsilon$, respectively. The evolution of bandgaps of TiO$_2$/GaSe heterostructure under a strain ranging from $-5\%$ to $\%5$ with a spacing of $1\%$ along zigzag and armchair directions is shown in Fig.\ref{fig:8}(a).

From Fig.\ref{fig:8}(a), we can see that as the lattice constants increase along the armchair direction, the bandgap decreases. However, as the lattice parameters augment along the zigzag direction, the bandgap increases. Meanwhile, for the strain along the armchair direction, the bandgap of the heterostructure is more sensitive to the tensile strain than compressive strain. On the contrary, when it comes to the zigzag direction, the bandgap changes more rapidly with compressive strain than tensile strain. In addition, the bandgap changes linearly when applying tensile strain along armchair direction or compressive strain along zigzag direction, which indicates such heterostructure may be developed as a mechanical sensor.

For better revealing the change of the band structure under biaxial strain, we studied in detail the variations of band edges of the heterostructure. As Fig.\ref{fig:8}(c) indicatess, along armchair direction, strain has more influence on the CBM than VBM, while strain almost equally affects the CBM and VBM along zigzag direction, which explains why bandgap changes more rapidly with strain along armchair direction than along zigzag direction. Meanwhile, with the increase of compressive strain along armchair direction, the CBM and VBM varies almost at the same gradient, which accounts for the insensibility for the bandgap to compressive along armchair direction. However, the inertia for bandgap to tensile strain along zigzag direction is due to the fact that tensile strain along zigzag direction nearly has no effect on the band edges of the heterostructure.

In the case of interlayer coupling, the bandgap and band edges as a function of the interlayer distance $d_i$ is shown in Fig.\ref{fig:8}(c, d). A lower $d_i$ than the equilibrium value denotes a compressive strain along the direction perpendicular to the surface, while a larger interlayer than the optimized value stands for a tensile strain. It indicates that as $d_i$ increases, the bandgap of the heterostructure first decreases rapidly, and then reaches a threshold of around 0.95eV. This is for the fact that the CBM of the heterostructure hardly changes with different $d_i$, while the VBM gradually increases with the increase of $d_i$ . Comparing Fig.\ref{fig:8}(a) with Fig.\ref{fig:8}(b), it can also be seen that to changing bandgap with biaxial strain is more effective than interlayer coupling. 

Such phenomena are mainly due to the fact that the applied tensile and compressive strain changes the distance between the atoms, leading to a different superposition of the atomic orbitals which result in the shift of the energy of the states.

\begin{figure*}[hbtp]
\centering
\includegraphics[scale=0.6]{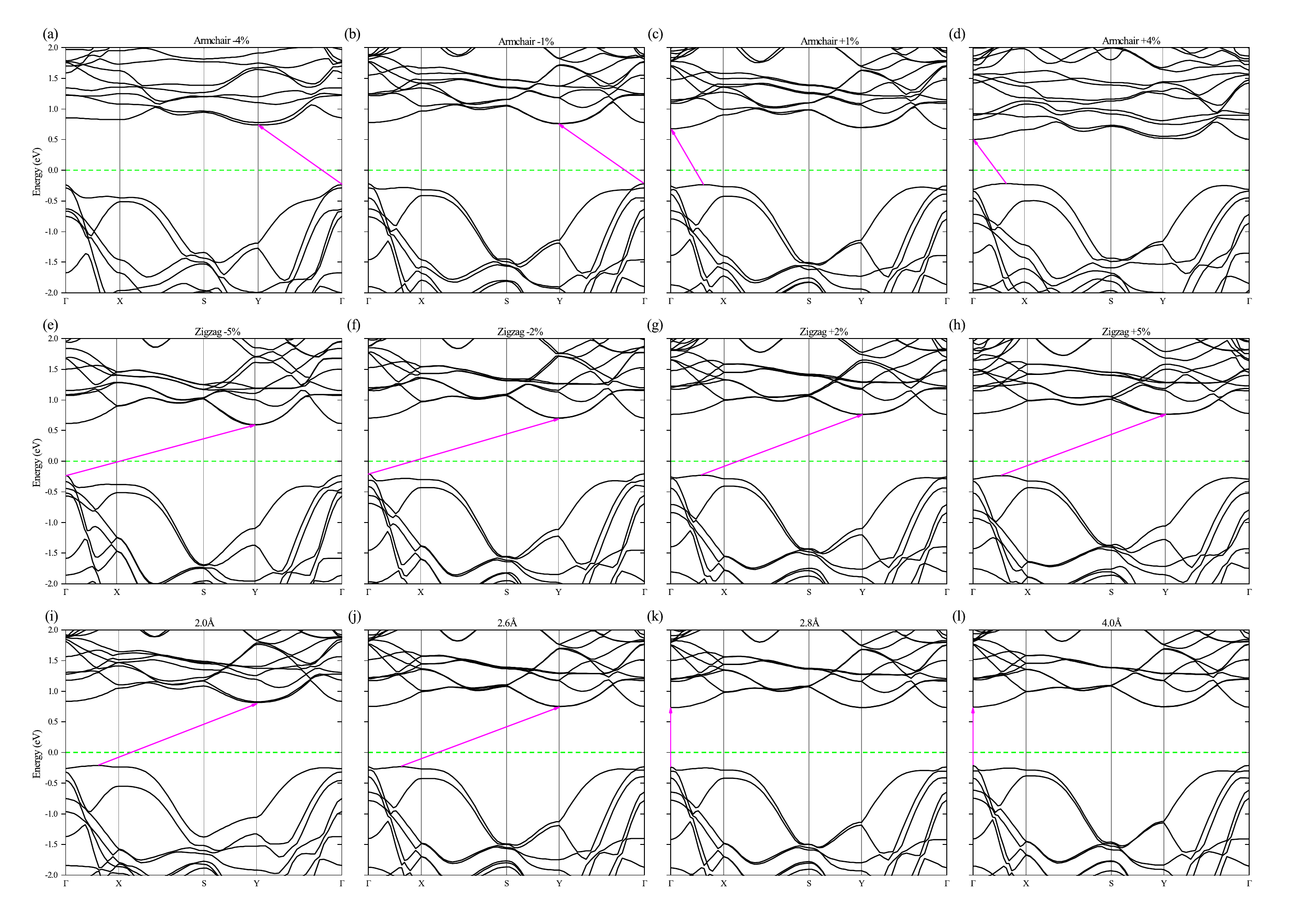}
\caption{\label{fig:9} Band structures of the TiO$_2$/GaSe heterostructure with (a-d) $\varepsilon=\pm1\%$, $\pm4\%$ along the armchair direction and (e-h) $\varepsilon=\pm2\%$, $\pm5\%$ along the zigzag direction. (i-l) Band structures of the TiO$_2$/GaSe heterostructure with interlayer distance $d_i$=2.0Å, 2.6Å, 2.8Å, 4.0Å, respectively.}
\end{figure*}

\begin{figure*}[hbtp]
\centering
\includegraphics[scale=0.6]{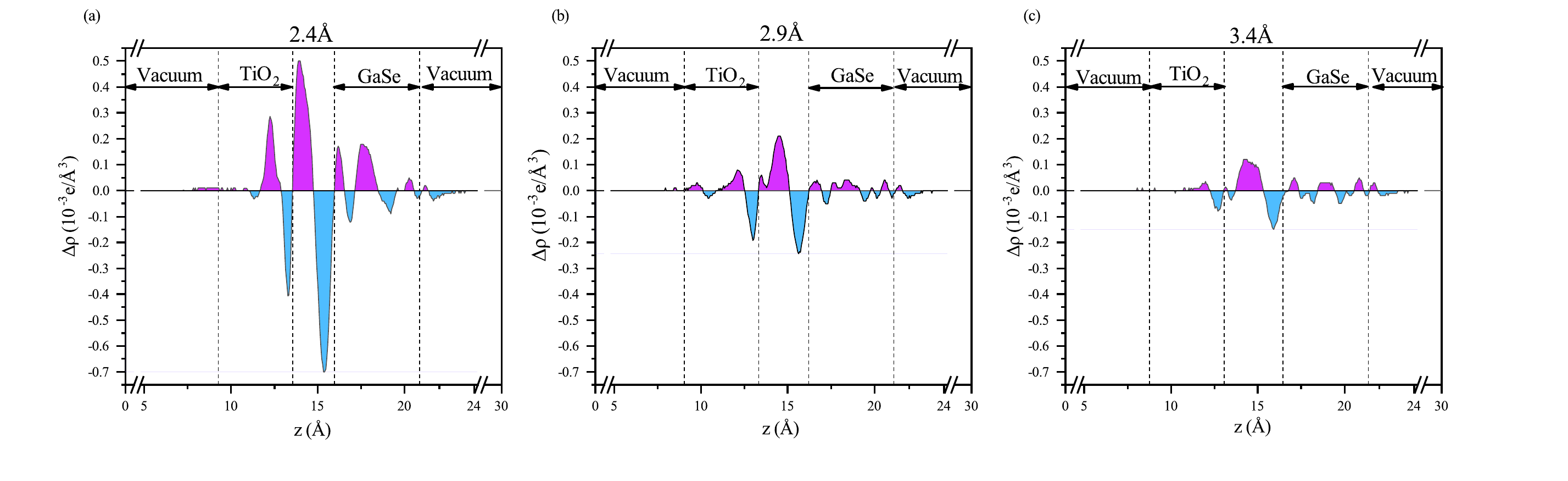}
\caption{\label{fig:10} The planar-averaged electron density difference $\Delta\rho(z)$ along z direction of the heterostructure with the interlayer distance di=2.4Å, 2.9Å and 3.4Å, respectively.}
\end{figure*}

Then, we further investigated the detailed band structures with different strains, as shown in Fig.\ref{fig:9}(a-l). It exhibits an obvious transition for the heterostructure from direct bandgap to indirect bandgap in all cases. The bandgap becomes indirect as long as the biaxial strain is applied, while for the interlayer coupling, the transition occurred when $d_i$ is less than 2.6Å. For biaxial strain, comparing with the optimized heterostructure, the compressive strain along the armchair direction drives the CBM from $\Gamma$ to Y. On the contrary, the tensile along the armchair direction cause the VBM to shift along $\Gamma$X line. Similar phenomena are also found when applying strain along the zigzag direction. For interlayer coupling, when the interlayer distance $d_i$ varies from 2.0Å to 2.6Å, the CBM of the heterostructure locates at the Y point and the VBM lies at the point on $\Gamma$X line. At the range of 2.8Å to 4.0Å, the CBM and VBM both lie at the $\Gamma$ point.

Essentially, such phenomena are mainly due to the fact that for the equilibrium heterostructure, the energy difference between the conduction band at the $\Gamma$ and Y is so small that a slight change in the distance between the atoms under the strain is sufficient to alter the energy states, leading to a transformation from direct bandgap to indirect bandgap. This is also true for the valence band at $\Gamma$ and the point on $\Gamma$X line.

Finally, we investigated how interlayer distance affects the charge redistribution. The Bader charge analysis indicates that about 0.193 electrons transfer from GaSe monolayer to TiO$_2$ surface when $d_i$=2.4Å. When $d_i$=3.4Å, the transferred electron number is 0.054, while for the optimized structure ($d_i$=2.9Å), the value was 0.099. The planar-averaged electron density difference of the heterostructure with different $d_i$ are plotted in Fig.\ref{fig:10}(a-c). As the figures suggest, with the decrease of the interlayer distance, the amount of the accumulated net charge on the interface increases, which manifests that charge redistribution can be enhanced by decreasing interlayer distance. This is for the fact that as the interlayer distance decrease, the interaction between the monolayers will increase, enhancing the charge redistribution at the interface.

\subsection{Ultrafast laser induced semiconductor-metal transition in TiO$_2$/GaSe heterostructure}
\label{sec:3.5}
\begin{figure}[h]
\centering
\includegraphics[scale=0.3]{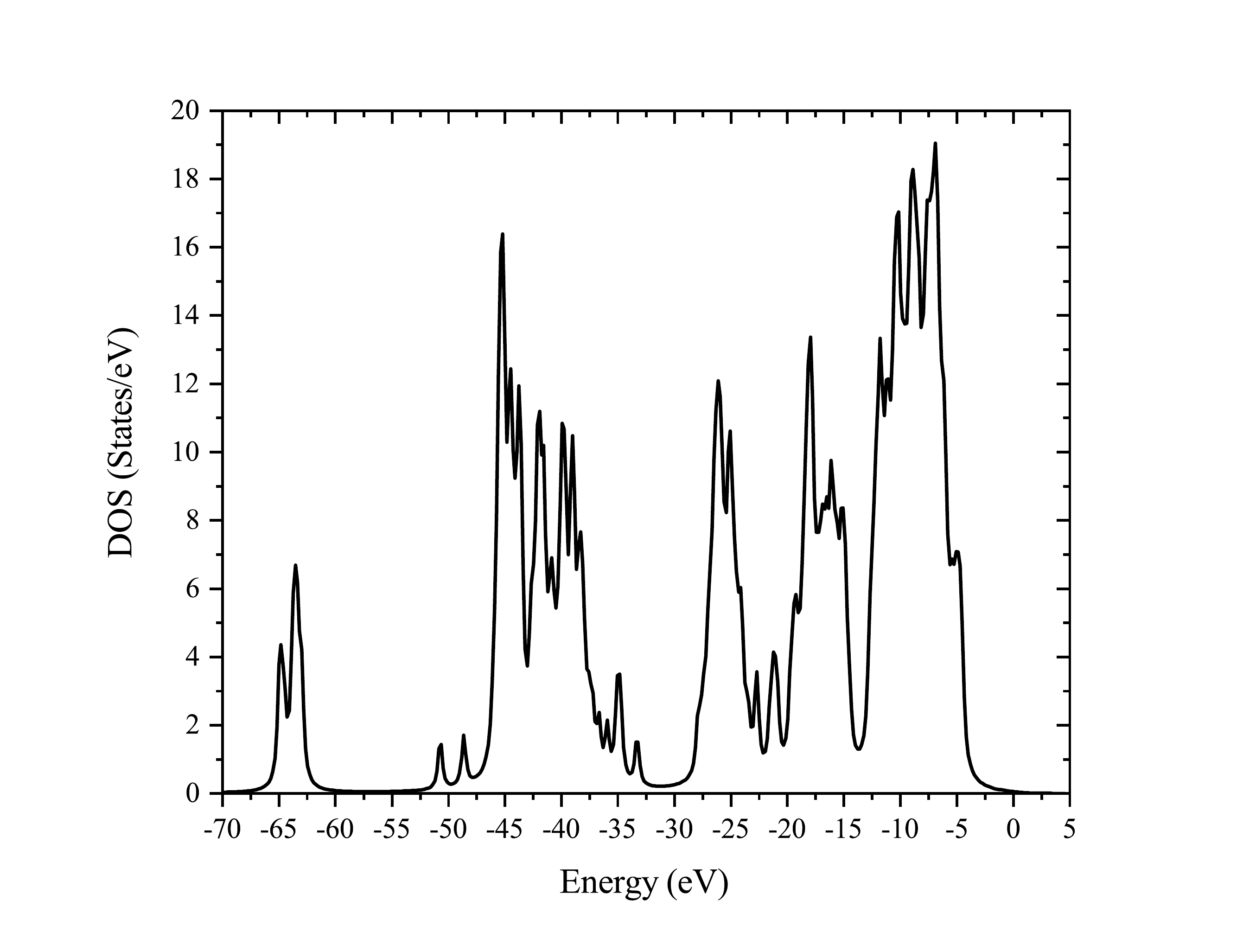}
\caption{\label{fig:11} Intrinsic density of states (DOS) of the TiO$_2$/GaSe heterostructure.}
\end{figure}

To explore the potential application for the TiO$_2$/GaSe heterostructure in developing optoelectronic devices, we finally calculated the electronic properties of the heterostructure under the illumination of ultrafast lasers with different intensities and different frequencies, respectively. The Fermi level is assumed to be determined by the external voltage bias and is immune to the electronic dynamics in the heterostructure. By investigating the intrinsic density of states (DOS) of TiO$_2$/GaSe, as shown in Fig.\ref{fig:11}, the change of the electron occupied states is characterized by the highest electron state (HES), as well as the number of electrons across the Fermi level ($N_F$). 

\subsubsection{Ultrafast laser with different intensity irradiating on the TiO$_2$/GaSe heterostructure}
\label{sec:3.5.1}

\begin{figure*}[htbp]
\centering
\includegraphics[scale=0.6]{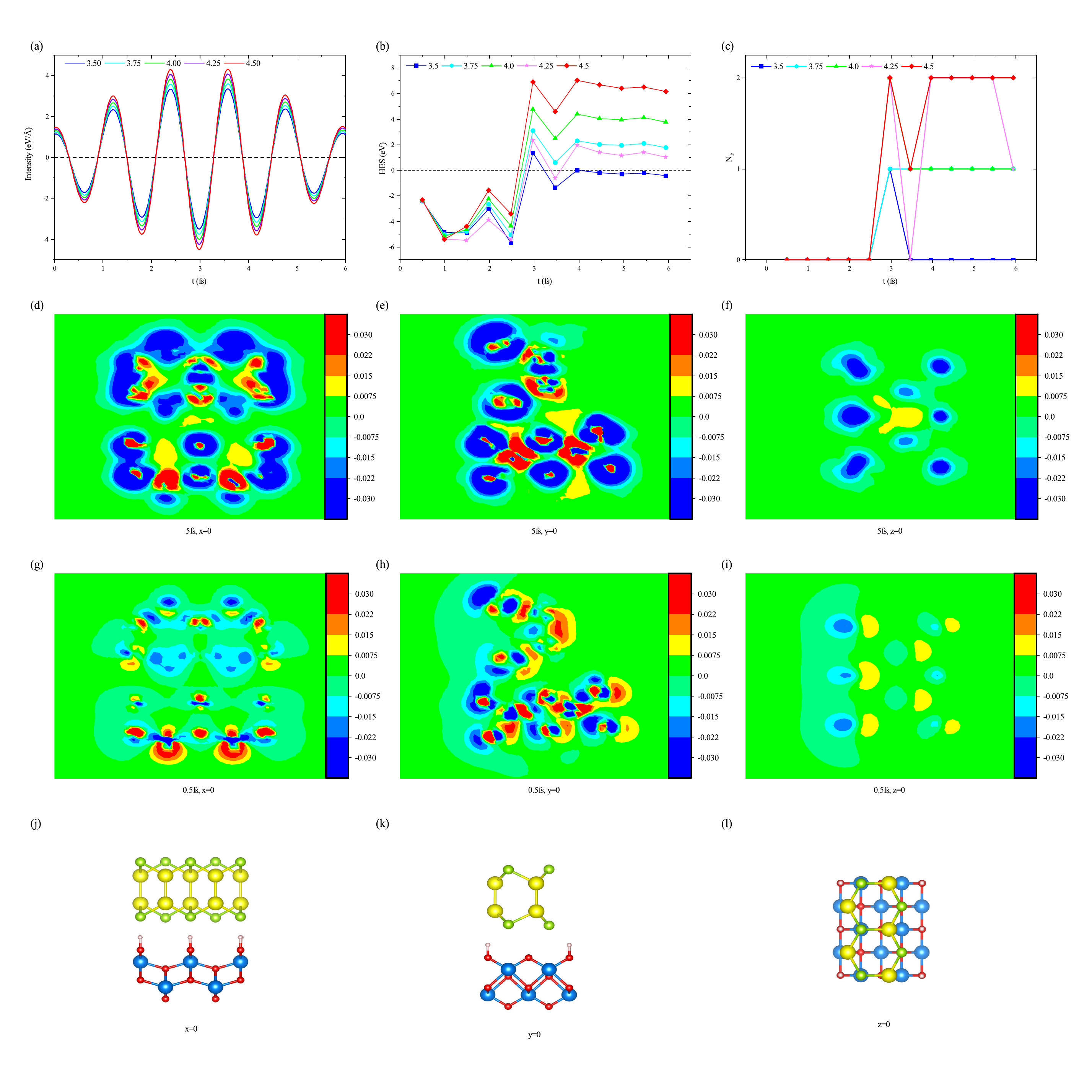}
\caption{\label{fig:12}  (a) Laser oscillogram. (b) The highest electron state (HES) and (c) the number of electrons across the Fermi level of TiO$_2$/GaSe as a function of time. Induced charge density distribution of TiO$_2$/GaSe at (d-f) 5fs and (g-i) 0.5fs with x-polarized laser intensity equals to 4.25 eV/Å. Electron accumulation and depletion are denoted as red and blue, respectively. (j) Front view, (k) side view and (l)top view of TiO$_2$/GaSe corresponding to induced charge density distribution. }
\end{figure*}

For the study of intensity of the ultrafast laser’s influence on the heterostructure, we studied two cases. In the first case, the intensity of the incident x-axis polarized laser with 358nm wavelength alter from 3.5 eV/Å to 4.5 eV/Å, with a spacing of 0.25 eV/Å. As shown in Fig.\ref{fig:12}(b, c), at first, the HES of the heterostructure just fluctuates with the propagation of the laser pulse. At 3fs, for the whole range of laser intensity, the HES starts to cross the Fermi level, which indicates a transition from semiconductor to metal. In addition, after 4fs, the HES stops fluctuates with the laser pulse and maintains the metallic electronic state afterwards. Why does such phenomenon occur? As aforementioned, the heterostructure can enhance the charge redistribution at the interface, and at equilibrium, the electrons diffusion is balanced with the built-in electric field which arises from the net charge accumulation. When the intense ultrafast laser was applied, this balance will be temporarily broken. By absorbing energy from the incident laser, more and more electrons are excited, enhancing the diffusion of electrons and holes, as well as the interaction between monolayers, which essentially leads to the semiconductor-metal transition. Meanwhile, with the increase of the number of accumulated charges at the interface, the built-in electric field gradually augments, too. However, since the laser varies very fast, the new balance between the built-in electric field and the diffusion of excitons cannot be achieved temporarily. As a result, the highest electron states (HES) fluctuate as the laser intensity changes, corresponding to the trend of HES in Fig.\ref{fig:12}(b) from 0fs to 4fs. Nevertheless, the intensity of the laser is damping. After 4fs, the value of the intensity is not more than 1eV/Å. With such a low intensity, no more electrons will be excited from the heterostructure. Consequently, the new equilibrium between the built-in electric field and the diffusion of excitons is reached, which explains why the HES still maintains the relatively high energy level and the heterostructure maintains the metal state since 4fs.

To further elucidate the charge redistribution in the process, we calculated the photo-induced charge density distribution in real-time propagation, as exhibited in Fig.\ref{fig:12}(d-i). It suggests that at 0.5fs, the induced electron density and the hole density are still localized under the weaker laser intensity, which manifests that the heterostructure is still a semiconductor. On the contrary, at 5fs, the more enhanced induced electron density and hole density becomes continuous and forms photo-induced currents, which suggests a semiconductor-metal transition.

\begin{figure*}[htbp]
\centering
\includegraphics[scale=0.6]{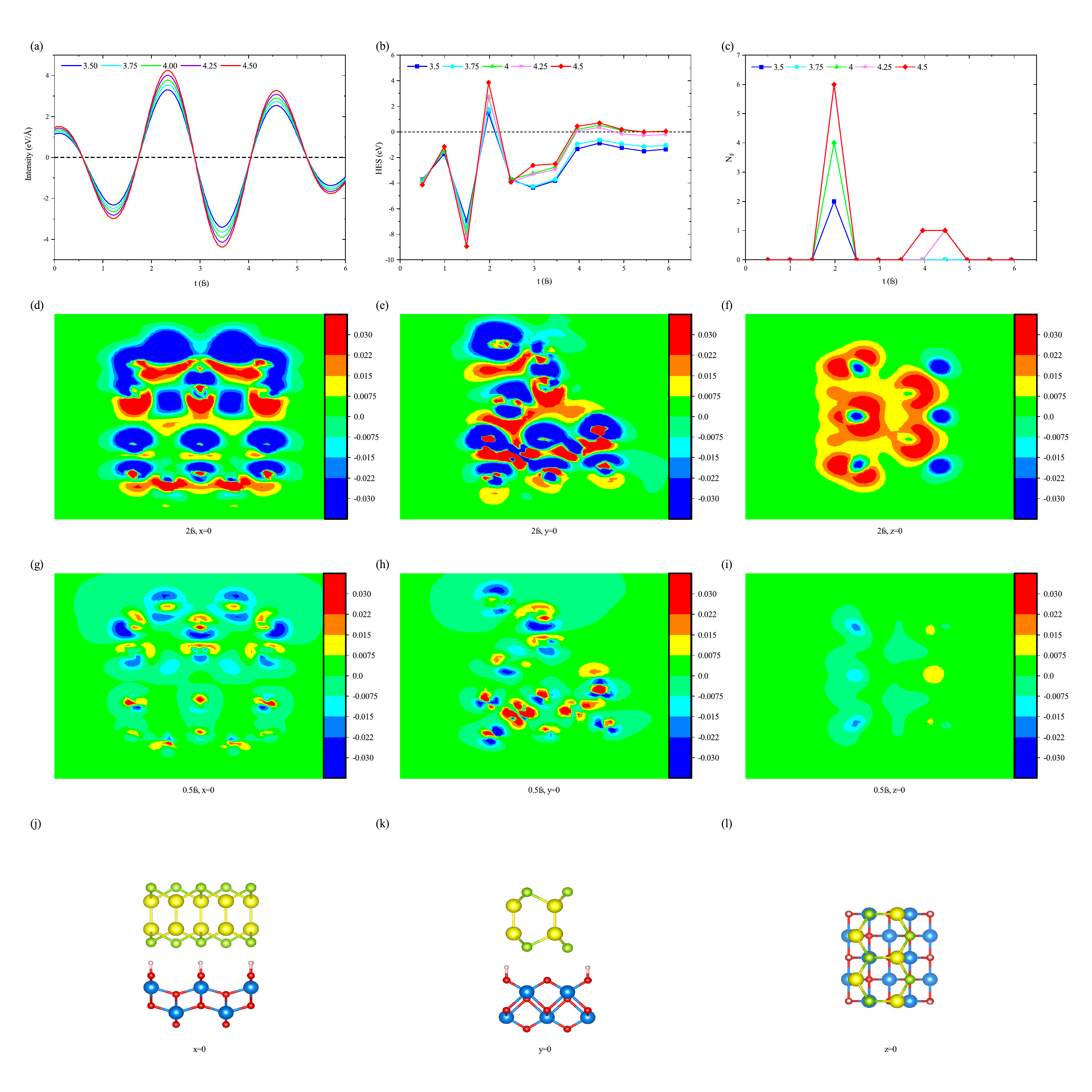}
\caption{\label{fig:13}   (a) Laser oscillogram. (b) HES and (c) $N_F$ of TiO$_2$/GaSe as a function of time. Induced charge density distribution of TiO$_2$/GaSe at (d-f) 2fs and (g-i) 0.5fs with z-polarized laser intensity equals to 4.25 eV/Å. Electron accumulation and depletion are denoted as red and blue, respectively. (j) Front view, (k) side view and (l)top view of TiO$_2$/GaSe corresponding to induced charge density distribution.}
\end{figure*}

In the other case, the incident laser was chosen to be z-axis polarized with 694nm wavelength, and the intensity also varies from 3.5 eV/Å to 4.5 eV/Å, with a spacing of 0.25 eV/Å. Similar phenomena have been found for the trend of HES and $N_F$ from Fig.\ref{fig:13}(b, c). The different point is that under high intensity laser, the photo-induced current is more intense than that in the case where laser is x-polarized, as shown in Fig.\ref{fig:13}(d-i), which means that the heterostructure is more sensitive to the laser along z direction. In addition, the photo-induced current is mainly at the interface of the heterostructure, which corresponds to the aforementioned conclusion that the heterostructure can enhance the charge redistribution at the interface, which causes the polarization of electrons and holes, leading to the formation of interface dipole. The behaviors of the plasmon resonance manifests that the coupling of TiO$_2$/GaSe and ultrafast laser has potential application in nanoscale plasmonic devices.

\subsubsection{Ultrafast laser with different frequency acting on the TiO$_2$/GaSe heterostructure}
\label{sec:3.5.2}
\begin{figure*}[htbp]
\centering
\includegraphics[scale=0.6]{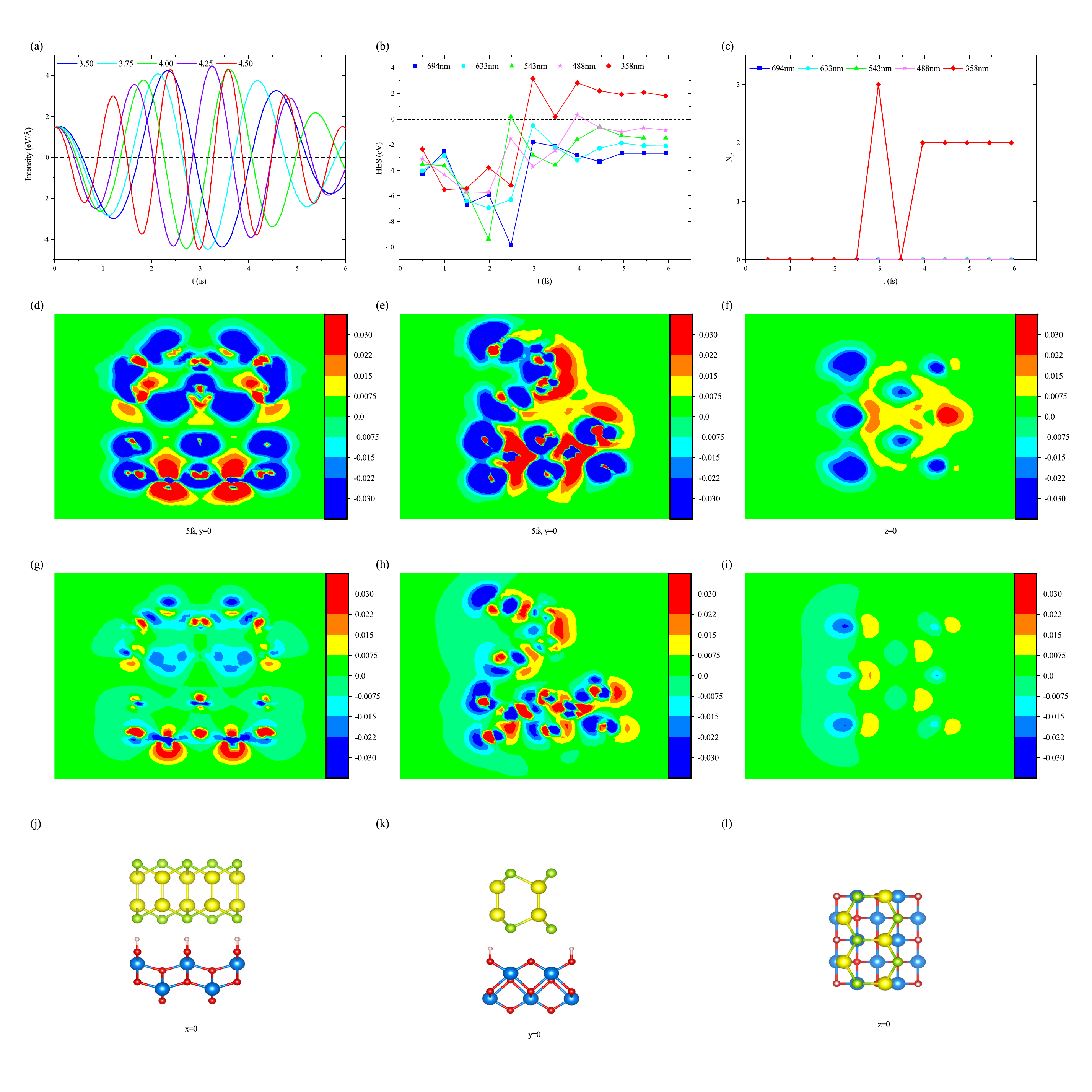}
\caption{\label{fig:14}    (a) Laser oscillogram. (b) HES and (c) $N_F$ of TiO$_2$/GaSe as a function of time. Induced charge density distribution of TiO$_2$/GaSe at (d-f) 5fs and (g-i) 0.5fs with x-polarized laser frequency equals to 358nm. Electron accumulation and depletion are denoted as red and blue, respectively. (j) Front view, (k) side view and (l)top view of TiO$_2$/GaSe corresponding to induced charge density distribution.}
\end{figure*}

\begin{figure*}[htbp]
\centering
\includegraphics[scale=0.6]{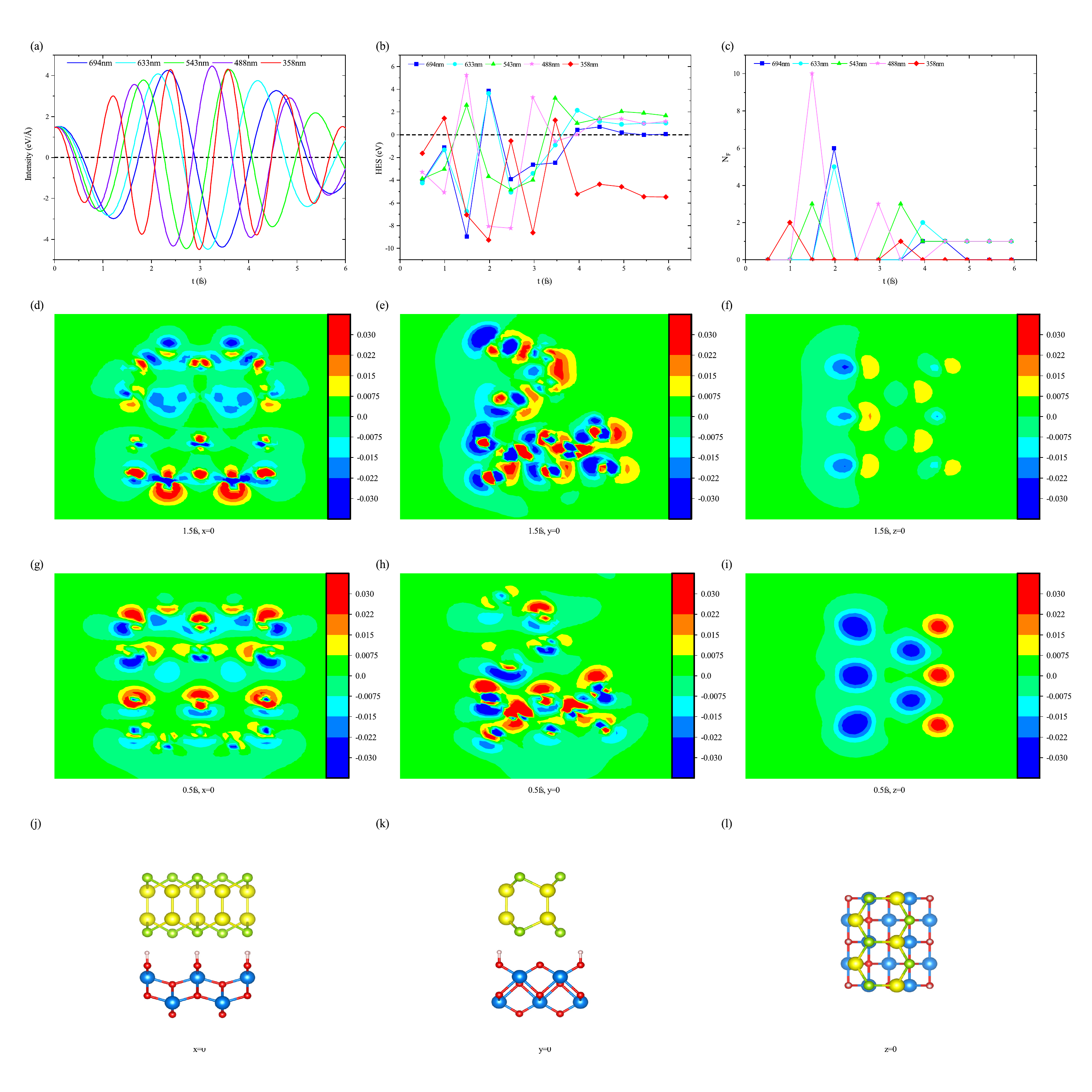}
\caption{\label{fig:15}    (a) Laser oscillogram. (b) HES and (c) $N_F$ of TiO$_2$/GaSe as a function of time. Induced charge density distribution of TiO$_2$/GaSe at (d-f) 1.5fs and (g-i) 0.5fs with z-polarized laser frequency equals to 488nm. Electron accumulation and depletion are denoted as red and blue, respectively. (j) Front view, (k) side view and (l)top view of TiO$_2$/GaSe corresponding to induced charge density distribution.}
\end{figure*}

Various wavelength lasers have always been applied in investigating the optical properties of all kinds of materials. To simulate lasers acting on TiO$_2$/GaSe in experiment, the laser wavelengths are set to be 694nm, 633nm, 543nm, 488nm, 358nm, corresponding to the ruby laser, Helium-Neon gas laser (red), Helium-Neon gas laser (green), Argon ion laser and Nitrogen laser, respectively. The intensity of the laser is chosen to be 4.5 eV/Å. As shown in Fig.\ref{fig:14}(b, c), the trends of the HES and $N_F$ are similar to these we discussed in Sec.\ref{sec:3.5.1}. What is new is that although the laser with other wavelength can also raise the HES of the system, only 358nm wavelength laser can transfer the system from semiconductor to metal, which indicates that the semiconductor-metal transition can be induced by tuning the wavelength of the incident laser. One reason for explaining the transition is that the eigenfrequency of the TiO$_2$/GaSe heterostructure is close to 358nm, which makes the electrons gain more resonance energy. Another reason is that 358nm is in the UV region, which is stronger in energy than the others. Consequently, the electrons may absorb more energy and are more likely to be excited.

In the case of laser along z direction, as exhibited in Fig.\ref{fig:15}(b, c), the trend of HES and $N_F$ also resembles the aforementioned ones, but unlike the case of x-direction laser, all of the wavelength can induce a semiconductor-metal transition for the system. This is also indicates that the heterostructure is more sensitive to the z-polarized laser. Moreover, the distribution of induced charge density, as shown in Fig.\ref{fig:15}(d-i), suggests that under the ultrafst laser, the electrons and holes at the interface are polarized and the interface dipole is formatted.

\section{CONCLUSIONS}
\label{sec:4}
In summary, we have systematically studied the properties of the lepidocrocite-type TiO$_2$/GaSe heterostructure based on first-principle calculations. From the analysis of the band structure and PDOS as well as the optical absorption spectrum, we find that the TiO$_2$/GaSe has a direct bandgap at the value of 0.963eV. It exhibits a type-II band alignment, a strong and broad optical absorption, ranging from visible light to UV region, which is beneficial to photovoltaic and photoelectronic devices. The planar-averaged electron density and Bader charge analysis further confirm its advantage in enhancing the separation of the electrons and holes, which is benefitial to photocatalysis. Through the investigation of bandgap modulation with biaxial strain and interlayer coupling, we find a direct-bandgap to indirect-bandgap transition both for biaxial strain and interlayer coupling. The linear change of bandgap with biaxial strain manifests its potential application in mechanical sensors. Finally, the calculation of ultrafast laser acting on TiO$_2$/GaSe further indicates that laser with specific wavelength and intensity can induce a semiconductor-metal transition for the heterostructure. Furthermore, the enhanced induced plasmonic current implys that TiO$_2$/GaSe coupling with ultrafast laser may be a good candidate in plasmonic devices. Our calculations provide valuable guidance for applying the lepidocrocite-type TiO$_2$/GaSe heterostructure in photovoltaic devices, photoelectronic devices, photocatalysis, mechanical sensors and plasmonic devices.

\begin{acknowledgments}
We acknowledge financial support from the National Key R$\&$D Program of China (2017YFA0303600), the National Natural Science Foundation of China (Grants No. 11974253) and the Science Speciality Program of Sichuan University (Grant No. 2020SCUNL210). X.C. acknowledges financial support from National Natural Science Foundation of China (Grants No. 11774248).
\end{acknowledgments}

\bibliography{abc}

\end{document}